\documentclass[preprint]{aastex}
\usepackage{graphics}
\usepackage{url}
\usepackage[section]{placeins}

\begin{document}

\title{The Effects of Refraction on Transit Transmission Spectroscopy: Application to Earth-like Exoplanets}
\author{Amit Misra\altaffilmark{1,2,3}}
\affil{Box 351580, UW}
\affil{Seattle, WA 98195-1580}
\affil{Phone: 206-616-4549}
\affil{Fax: 206-685-0403}
\email{amit0@astro.washington.edu}
\and
\author{Victoria Meadows\altaffilmark{1,2,3} \& Dave Crisp\altaffilmark{4,2}}
\altaffiltext{1}{Box 351580, Univ. of Washington Astronomy Dept., Seattle, Washington, USA 98195}
\altaffiltext{2}{NAI Virtual Planetary Laboratory, Seattle, Washington, USA}
\altaffiltext{3}{Univ. of Washington Astrobiology Program, Seattle, Washington, USA}
\altaffiltext{4}{M/S 183-501, 4800 Oak Grove Drive Jet Propulsion Laboratory, California Institute of Technology, Pasadena, California, USA 91109}

\begin{abstract}

We quantify the effects of refraction in transit transmission spectroscopy on spectral absorption features and on temporal variations that could be used to obtain altitude-dependent spectra for planets orbiting stars of different stellar types. We validate our model against altitude-dependent transmission spectra of the Earth from ATMOS and against lunar eclipse spectra from \citet{palle09}. We perform detectability studies to show the potential effects of refraction on hypothetical observations of Earth analogs with the \textit{James Webb Space Telescope} NIRSPEC. Due to refraction, there will be a maximum tangent pressure level that can be probed during transit for each given planet-star system. We show that because of refraction, for an Earth-analog planet orbiting in the habitable zone of a Sun-like star only the top 0.3 bars of the atmosphere can be probed, leading to a decrease in the signal-to-noise ratio (S/N) of absorption features by 60\%, while for an Earth-analog planet orbiting in the habitable zone of an M5V star it is possible to probe almost the entire atmosphere with minimal decreases in S/N. We also show that refraction can result in temporal variations in the transit transmission spectrum which may provide a way to obtain altitude-dependent spectra of exoplanet atmospheres. Additionally, the variations prior to ingress and subsequent to egress provide a way to probe pressures greater than the maximum tangent pressure that can be probed during transit. Therefore, probing the maximum range of atmospheric altitudes, and in particular the near-surface environment of an Earth analog exoplanet, will require looking at out-of-transit refracted light in addition to the in-transit spectrum.
\end{abstract}

\section{Introduction} 

As an exoplanet transits its host star, some of the star's light is absorbed and scattered in the planet's atmosphere, and the resulting transit transmission spectrum can be used to characterize the planet's atmospheres \citep{seager00, brown01, hubbard01}. This technique has been used to characterize many Hot Jupiters \citep{char02, vidal03, pont08} and for some mini-Neptune/super-Earths \citep{kreidberg14, knutson14}. With the upcoming launch of the \textit{James Webb Space Telescope} (\textit{JWST}) and the construction of larger ground-based telescopes, it may be possible to use transit transmission spectroscopy to characterize terrestrial planets in the near future \citep{deming09, belu11, snellen13}.

Solar and stellar occultation measurements collected in our solar system can serve as analogs for exoplanet transit transmission observations. Occultation spectra can be analyzed by taking the ratio of the light transmitted through a planet's atmosphere and unattenuated starlight recorded prior to or after the occultation (see reviews by \citet{smith90, elliot96}). These observations have been studied for numerous planets and moons in the solar system. One strength of occultation measurements in the solar system is that they provide vertical sounding, meaning the observations can provide altitude-dependent atmospheric spectra. For example, recent occultation measurements by Mars Express and Venus Express have constrained altitude-dependent gas mixing ratios and aerosol profiles in the atmospheres of Mars and Venus (e.g., \citet{fedorova09, wilquet12}). In this paper, we will show how similar spectra could be obtained for exoplanet observations. 

As light traverses a planetary atmosphere, it is bent or ``refracted" in response to the atmospheric index of refraction gradient. This process is important for interpreting occultation observations because it can modify the atmospheric path length traversed by the light. Refraction has been shown to be significant for solar system observations, such as lunar eclipse observations, transits of Venus, and radio and stellar occultations. In lunar eclipse spectra, the light that reaches the eclipsed Moon is refracted by Earth's atmosphere. During the 2004 transit of Venus, a rim of refracted light could be seen on the trailing side of Venus before that part of the Venus disk crossed the solar limb \citep{pasachoff2011}. Finally, in planetary radio occultations, refraction limits the depth to which the radio beam from a spacecraft can penetrate the planet's atmosphere and yields information about the atmospheric density and thermal structure \citep{tyler82}. While radio occultations are not currently feasible for exoplanets, the principle of refraction will still apply to exoplanet transit transmission observations. 
 
The impact of refraction on transit transmission spectra has been previously examined by \citet{hui02, sidis10, garcia12}. \citet{hui02} and \citet{sidis10} explore the effect of refraction on an exoplanet light curve, noting that starlight will be deflected to a distant observer outside of transit due to refraction, as seen in the 2004 Venus transit. They also show that some starlight will be deflected away from a distant observer during a transit. This concept is examined further by \citet{garcia12}, who investigate the effect of refraction on lunar eclipse spectra and exoplanet transit transmission spectra. Here, we use a new line-by-line radiative transfer model that includes refraction to expand on these concepts by quantifying the detectable effects of refraction on transit light curves and spectra and by presenting a novel method for obtaining altitude-dependent spectra of exoplanets via refraction. We describe the model in Section \ref{sec:methods} and provide a model validation in Section \ref{sec:validation}. In Section \ref{sec:respressure}, we quantify the maximum tangent pressures and generate spectra for Earth-like planets orbiting stars of different spectral types. In Section \ref{sec:sounding}, we describe a new method to obtain altitude-dependent spectra of an exoplanet from temporal variations in the transit transmission spectrum due to refraction. In Section \ref{sec:detect}, we examine the detectability of refractive effects on the transit light curve and the transmission spectrum and show that refraction can reduce the detectability of biosignature gases.


\section{Methods} \label{sec:methods}

\subsection{Transmission Spectrum Model}

The transit transmission spectroscopy model combines spectrally dependent atmospheric optical properties obtained from the Spectral Mapping Atmospheric Radiative Transfer (SMART) model \citep{meadows96, crisp97} with a limb transmission model. The transmission model includes gas absorption, Rayleigh scattering, collision-induced and binary (dimer) absorption, extinction from clouds and aerosols, refraction and limb darkening. We describe the details of this model below. 

SMART calculates monochromatic optical properties in each layer of a vertically homogenous model atmosphere given the vertical profiles of pressure, temperature, gas mixing ratios, and aerosol optical depth. We used a program (Line By Line ABsorption Coefficients, or LBLABC) to calculate line-by-line absorption coefficients for all absorbing gases in the model atmosphere (H$_2$O, CO$_2$, O$_2$, N$_2$O, CH$_4$, CO, O$_2$) given a model atmosphere. For the calculations presented here, LBLABC uses the HITRAN 2008 line lists \citep{rothman09}. SMART reads in the line-by-line gas absorption cross-sections from LBLABC at each spectral grid point and atmospheric level and combines them with the gas mixing ratios and layer thickness to derive the normal-incidence (for a vertical path through the layer) gas extinction and optical depths. These values are combined with the corresponding extinction optical depths for Rayleigh scattering, clouds, and aerosols to yield the total normal-incidence extinction optical for each layer at each spectral point.

To generate transit transmission spectra, the normal-incidence optical depths calculated by SMART are used in a transit transmission model that describes limb-traversing trajectories. The path length between two adjacent layers $j$ and $j+1$ for a path $p(i,j)$ with a tangent altitude of $h(i)$ is
\begin{equation}
p(i,j) = \sqrt{h(j+1)^2-h(i)^2}-\sqrt{h(j)^2-h(i)^2} \label{eqn:path}
\end{equation}

The transit transmission spectrum for each tangent altitude can be generated by combining the opacities calculated by SMART and the path lengths as given in Equation (\ref{eqn:path}) using the Beer-Lambert law. A full transit transmission spectrum is the weighted average of the altitude-dependent spectra, with the weight dependent on the area of planetary annulus at each altitude.
\begin{eqnarray}
F(\lambda, i) &=& e^{- \sum_j n(j)*\sigma (\lambda)*p(i,j)} \\
F(\lambda) &=& \frac{\sum_i F(\lambda, i)*\pi[(h(i+1) + R_{P})^2 -(h(i)+ R_{P})^2]}{\pi (max(h) + R_{P})^2}
\end{eqnarray}
where $F(\lambda,i )$ is the percent of flux transmitted at each wavelength $\lambda$ for each altitude layer $i$, $\sigma(\lambda)$ is the absorption and extinction cross section of the atmosphere wavelength $\lambda$, and $n(j)$ is the number density at altitude layer $j$. $F(\lambda)$ is the percent of flux transmitted through the entire atmosphere, and $R_{P}$ is the radius of the planet.

\subsection{Refraction} \label{sec:refraction}

Light refracts or bends as it passes through a medium with an index of refraction gradient, such as a planetary atmosphere. As noted in \citet{sidis10}, the primary effect of refraction for exoplanet transits is that starlight can be refracted into the beam to a distant observer prior to a transit event and out of the beam to a distant observer during a transit.

We include refraction in our transit transmission spectroscopy model using a path integration method described in \citet{werf08}. Given an index of refraction profile and the initial trajectory of a ray we can calculate the full path of the ray through the atmosphere and the final angle at which it leaves the atmosphere. We used a step size of 10 km to trace the photon path through the atmosphere, though we find that smaller step sizes do not lead to different results. At each step along the path, we calculated the index of refraction at that point in the atmosphere by linearly interpolating between the layers of the model atmosphere. We then use a fourth-order Runge-Kutta integration method to solve a series of differential equations that determine the change in the path trajectory. Once a ray reaches the maximum altitude in the model atmosphere, we assume it has left the atmosphere and record the final trajectory. 

The final angle calculated from the path integration model can be used to determine if a ray at a particular tangent altitude in the atmosphere can connect a distant observer to the host star. For each altitude and each portion of the planetary annulus, we determine from where the path would need to have originated from to reach a distant observer. If this path originates at the stellar disk, then that portion of the annulus of the planetary atmosphere can be probed via transit transmission spectroscopy. If the path does not originate within the stellar disk, that portion of the annulus will appear opaque, independent of atmospheric opacity. Because different portions of the atmosphere can connect with different parts of the stellar disk during a transit, stellar limb darkening is included in our refractive transit transmission spectroscopy model. An accurate treatment of limb darkening is necessary to correctly determine the total amount of flux transmitted through the planet's atmosphere. We used the limb darkening coefficients of \citet{claret00} for the present work, as they are valid over a range of stellar effective temperatures, metallicities, and surface gravities, and so are applicable to a range of stellar types. 

\subsection{Model Input Parameters} 

\subsubsection{Solar and Stellar Parameters}

We modeled the effects of refraction over a range of stellar types from G2V to M9V, using the stellar parameters found in \citet{reid05} for the M dwarfs and \citet{zombeck90} for the earlier-type stars. For each star, we placed an Earth analog planet at the flux-equivalent distance, that is, the distance from the star at which it receives the same bolometric flux as the Earth does from the Sun today. We highlighted two test cases for our analysis of the effects of refraction and detectability studies: the Earth analog planet orbiting a Sun-like star and M5V star. For the solar spectrum, we used a Kurucz model atmosphere \citep{kurucz79}. The other stellar spectra are Phoenix NextGen \citep{hauschildt99} model spectra. In particular, the M5V spectrum has a temperature of 2800 K, a log of surface gravity of 5.0, and solar metallicity.


\subsubsection{Model Atmospheres}

Model atmospheres for Earth-like planets were developed to validate the model and to simulate Earth-like exoplanets. These models describe the temperature, trace gas mixing ratio, and aerosol profiles as a function of pressure. 

The model atmospheres used for the validation experiments were customized to simulate the conditions observed at the time the validation data were recorded. Two validation experiments were performed. The first used solar occultation spectra collected by the Atmospheric Trace Molecule Spectroscopy (ATMOS) Experiment \citep{gunson90}. ATMOS was designed to determine the volume mixing ratio profiles of minor and trace gases in Earth's atmosphere. This was done by obtaining numerous solar occultation spectra through the limb of Earth's atmosphere at wavelengths between 2 and 17 $\mu$m and by using a global retrieval algorithm to simultaneously retrieve volume mixing ratio profiles at all altitudes within each occultation \citep{irion02}. The second validation source used is lunar eclipse observations \citep{palle09}. During a lunar eclipse, the sunlight that reaches the Moon's surface traverses Earth's atmosphere on limb trajectories and is refracted, producing a spectrum similar to a transit transmission spectrum.

\begin{figure}
\centering
\includegraphics[width=8cm]{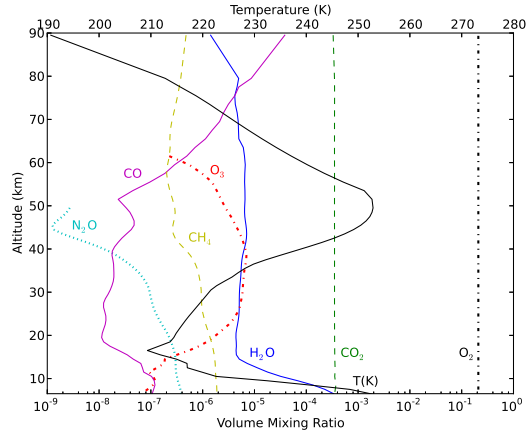}
\caption{Temperature profile and gas volume mixing ratios used to generate model ATMOS spectra. The profiles are the averages of the ATMOS retrievals between 40 and 50$^{\circ}$ N. \label{fig:atmos-input1}}
\end{figure}

\begin{figure}
\centering
\includegraphics[width=8cm]{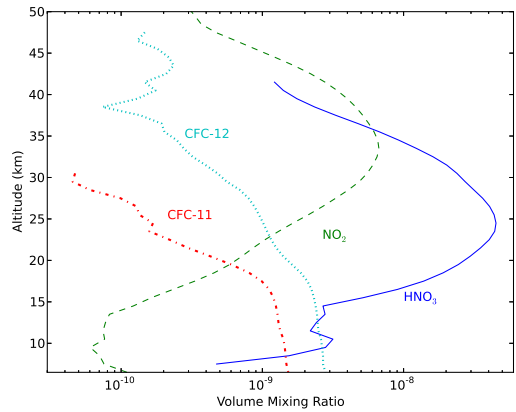}
\caption{Average retrieved gas volume mixing ratios from ATMOS for HNO$_3$, NO$_2$, CFC-11, and CFC-1. These gases are predominantly from anthropogenic sources. \label{fig:atmos-input2}}
\end{figure}

To simulate synthetic spectra to match the ATMOS data, we used the ATMOS retrieved altitude-dependent gas mixing ratios and temperature profile. We used the average of the retrieved abundances of H$_2$O, CO$_2$, O$_3$, CO, N$_2$O, CH$_4$, O$_2$, NO$_2$, HNO$_3$, CFC-11, and CFC-12 for ATMOS occultations taken at latitudes between 40$^{\circ}$ and 50$^{\circ}$N. We used N$_2$-N$_2$ and N$_2$-O$_2$ absorption cross-sections to model absorption near 4.3 $\mu$m (\citet{lafferty96}; Eddie Schwieterman, private communication). Figure \ref{fig:atmos-input1} shows the volume mixing ratio profiles for many of the gases included in the model runs as well the temperature profile. Figure \ref{fig:atmos-input2} shows the volume mixing ratio profiles for NO$_2$, HNO$_3$, CFC-11 and CFC-12, which are primarily anthropogenic. We generated high-resolution spectrum over the entire wavelength range, then convolved it with a triangle slit to a spectral resolving power of \textit{R}=1000 when comparing the model spectra to the ATMOS data.

\begin{figure}
\centering
\includegraphics[width=8cm]{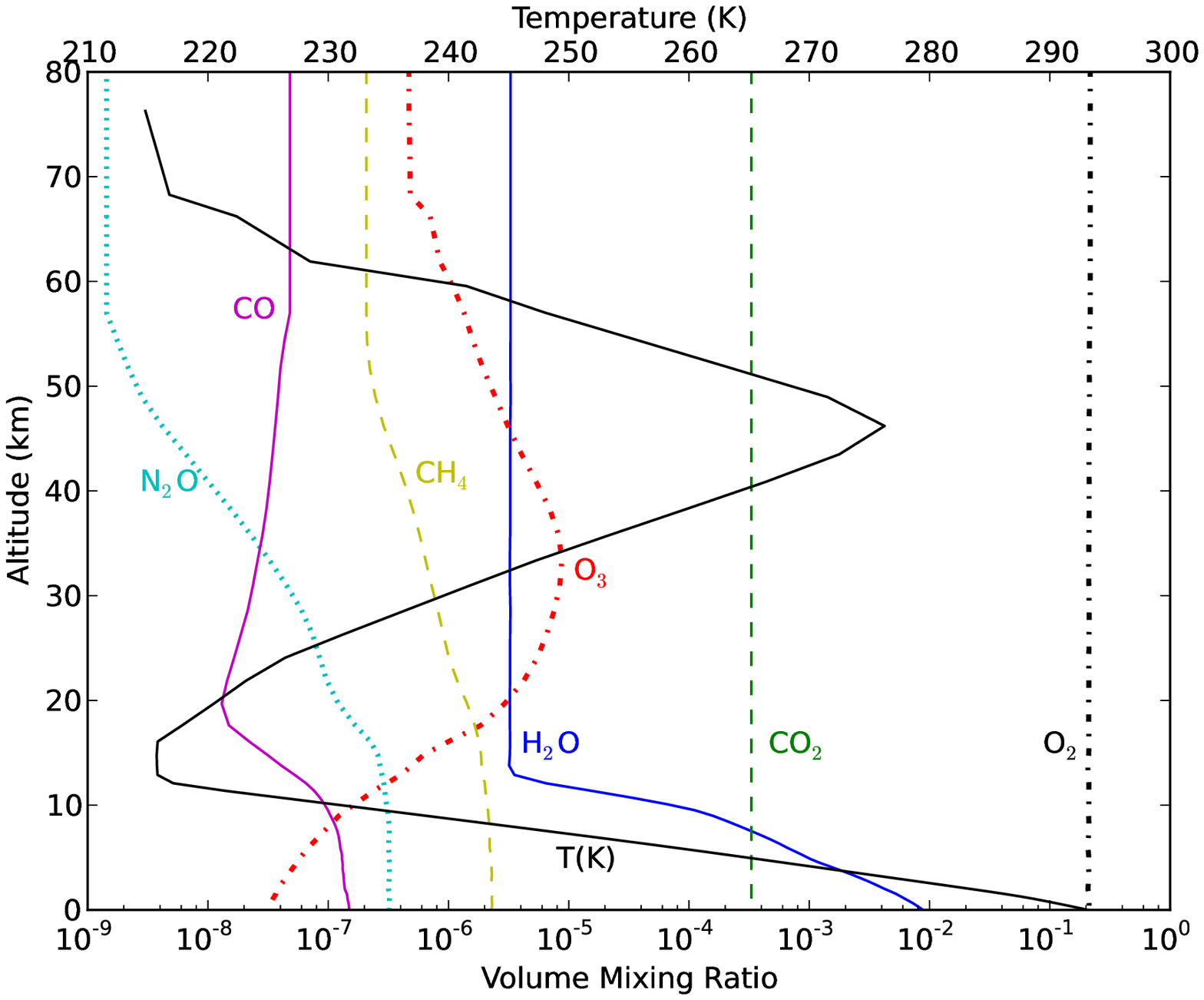}
\caption{Temperature and mixing ratio profiles from the U.S. standard summer northern mid-latitudes model atmosphere. We used this model atmosphere to generate lunar eclipse spectra and Earth analog spectra. We note that for the lunar eclipse spectra we scaled down the H$_2$O mixing ratio by a factor of two.\label{fig:earth-lunar-eclipse}}
\end{figure}

The model atmosphere developed for the lunar eclipse validation was based on the mid-latitude summer model atmosphere adopted for the Intercomparison of Radiation Codes in Climate Models (ICRCCM) \citep{mcclatchey72, clough92}. The temperature profile and volume mixing ratio profiles for this model atmosphere are shown in Figure \ref{fig:earth-lunar-eclipse}. To produce an acceptable fit to the lunar eclipse spectra, which predominantly probe latitudes greater than 50$^{\circ}$, we decreased the H$_2$O mixing ratio by a factor of two from the standard ICRCCM atmospheric profile to more accurately depict the expected H$_2$O mixing ratio profiles at those latitudes. Following the method of \citet{garcia12}, we also assume any path with a tangent altitude of 6 km or less is opaque because of clouds. In addition to the vibration-rotation transitions for the the gases shown in Figure \ref{fig:earth-lunar-eclipse}, we also included the ``interaction-induced absorption" associated with collision-induced absorption bands and dimers for O$_2$-O$_2$ and O$_2$-N$_2$. We used the O$_2$-O$_2$ cross-section from \citet{greenblatt90} and the O$_2$-N$_2$ cross-section from \citet{mate99} to model absorption at 1.06 $\mu$m and 1.27 $\mu$m. As with the ATMOS spectra, we initially generated high-resolution spectra, and then we convolved it with a triangle slit to match the resolving power of the lunar eclipse data, which was at \textit{R}$\sim$1000.

We also include aerosol extinction in the lunar eclipse model atmosphere by using data from \citet{sioris10}. They use the \textit{Atmospheric Chemistry Experiment} (ACE) on board Measurements of Aerosol Extinction in the Stratosphere and Troposphere Retrieved by Occultation to determine the amount and size distribution of aerosol particles in the atmosphere. We used their data from 2007 September to model the lunar eclipse data because by 2008 September, the aerosol levels had dramatically increased due to the Kasatochi eruption in 2008 August. The lunar eclipse occurred nine days after the eruption, which was not enough time for volcanic aerosols to be detectable \citep{bourassa10}. Additionally, we used a log-normal aerosol size distribution with a mean of 0.5 $\mu$m and a variance of 0.28$\mu$m$^2$, which we found to provide the best fit to the data and are consistent with the results from \citet{sioris10}. However, we note that changing the mean to 0.4 $\mu$m or 0.6 $\mu$m and letting the variance change by over an order of magnitude did not greatly increase our model validation errors. Our aerosol cross-sections were calculated using a python version of the code \texttt{bhmie} \citep{bohren83}.\footnote{Retrieved from Ray Pierrehumbert's ``Principles of Planetary Climate'' Web site.}

The model atmosphere developed for simulating Earth-like exoplanets is the ICRCCM model atmosphere, and includes absorption from the gases shown in Figure \ref{fig:earth-lunar-eclipse} along with O$_2$-O$_2$, O$_2$-N$_2$ and N$_2$-N$_2$ absorption. The exoplanet model atmosphere does not include clouds or aerosol particles. We convolved our high resolution spectra to a resolving power of R=100, applicable to the \textit{JWST} NIRSPEC (Near Infrared Spectrograph) instrument in single prism mode \citep{kohler05}.

\subsection{Temporal Variations} 

With our model, we examined temporal variations in a transit transmission spectrum due to refraction. As a planet progresses through its transit the exoplanet's position relative to the stellar disk will change. Therefore, different portions of the annulus of the planetary atmosphere will be backlit (meaning light is transmitted through the atmosphere to a distant observer), leading to variations in the transit transmission spectrum of the exoplanet. For each altitude in a model atmosphere, we used the method outlined in Section \ref{sec:refraction} to determine the initial trajectory of light that traverses the atmosphere and is capable of reaching a distant observer. Whether or not this trajectory connects with the stellar disk depends on the location on the annulus of the planetary atmosphere. For each altitude and position on the annulus of the planetary atmosphere, we calculate where the initial trajectory connects with the stellar disk throughout the course of a transit event. This allows us to calculate the flux through each portion of the atmosphere. We can generate a temporally varying transit transmission spectrum from this by summing over all altitudes and positions on the planetary atmosphere's annulus at each stage during the transit event.

\subsection{Error Analysis} \label{sec:erroranalysis}

To quantify the discrepancies between model and data spectra we calculated rms errors. We define the absolute rms error as the rms of the difference between the two spectra. We define the relative rms error as the rms of the difference divided by the data. 

We performed detectability studies \citep{kalt09, deming09, belu11, rauer11} for an Earth analog orbiting stars ranging from Sun-like (G2V) to an M9V star, with and without refraction included. We followed the detectability calculation method described in \citet{misra14}. We used the \textit{JWST} Exposure Time Calculator (ETC) \citep{sosey12}\footnote{http://jwstetc.stsci.edu} to calculate part per million (ppm) noise levels at each wavelength, assuming all transits over \textit{JWST}'s five-year mission lifetime could be observed and coadded. The noise is dominated by photon noise in all cases considered. The signal is the ppm change in flux integrated over the entire absorption band. The final signal-to-noise ratio (S/N) is the absorption signal divided by the noise calculated from the \textit{JWST} ETC, divided by $\sqrt{2}$. The factor of $\sqrt{2}$ must be included because the total noise will be the sum of the squares of the noise for the in-transit and out-of-transit spectrum. We assume the in-transit and out-of-transit noise levels are equal, so the S/N must be divided by $\sqrt{2}$. For the temporal variations spectra, we divide the signal by the noise and a factor of $\sqrt{3}$ because we have divided the transit into three different stages, and therefore the total integration time for each spectrum is decreased by three, leading to an increase in noise by a factor of $\sqrt{3}$. 


\section{Results}

\subsection{Model Validation Results} \label{sec:validation}

\subsubsection{ATMOS Validation}


We used the ATMOS sunset spectra from the ATLAS 3 space shuttle with latitudes between 40$^{\circ}$ and 50$^{\circ}$ taken between 1994 November 4 and 6\footnote{Data available at http://remus.jpl.nasa.gov/atmos/atmos.html}. The retrieved atmospheric properties were taken from the ATMOS version 3 retrievals described in \citet{irion02}. The average signal to noise varies by spectral filter, with values of 74$\pm$11 for Filter 3 (2.9-6.3 $\mu$m), 98$\pm$35 for Filter 4 (2.1-3.2 $\mu$m), 122$\pm$40 for Filter 9 (4.1-16.7 $\mu$m), and 255$\pm$36 for Filter 12 (7.1-16.7 $\mu$m). The spectra were measured with a spectral resolution of $\sim$0.01 cm$^{-1}$ (\textit{R}$\sim$10$^6$ at 1.0 $\mu$m), but we applied a triangle filter to the spectra with a resolving power of 1000 when comparing the spectra to our model spectra.

\begin{figure*}
\centering
\includegraphics[width=16.5cm]{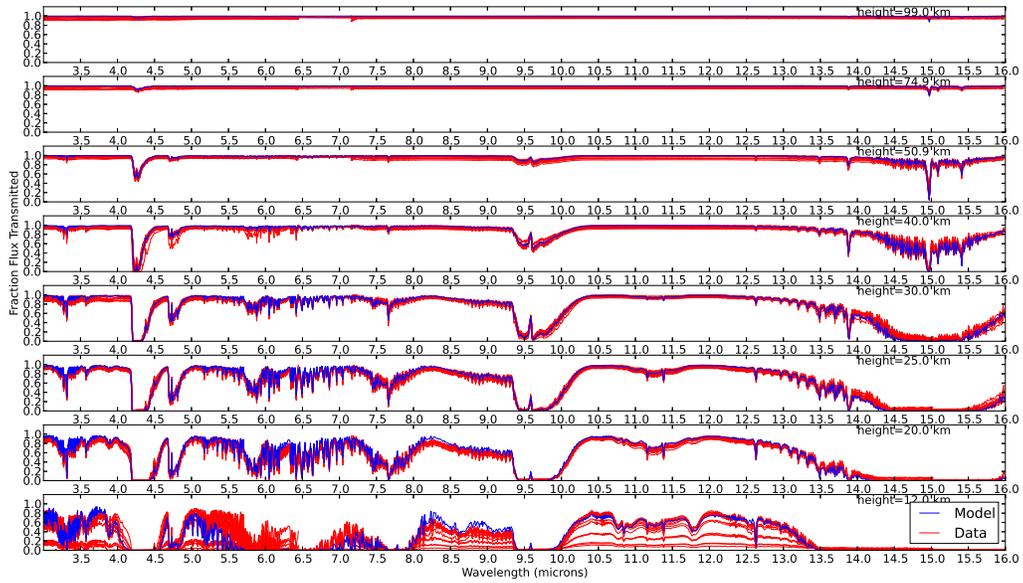}
\caption{Altitude-dependent ATMOS spectra (red) and model spectra (blue). Multiple ATMOS spectra are shown for each wavelength band and altitude. The model and data agree very well except at altitudes of 12 km and below. At these altitudes, clouds become important and vary spatially and temporally, leading to a large amount of variation in the 12 km ATMOS data. \label{fig:atmos}}
\end{figure*}

Figure \ref{fig:atmos} shows a comparison of the ATMOS data with our model spectra. ATMOS data were taken over a wider range of altitudes than shown here, and the model can produce spectra at any given altitude; however, these results were shown for simplicity and consistency with previous work \citep{kalt09}. Multiple data spectra are shown for each altitude. Table \ref{tab:atmos} shows the discrepancies between the ATMOS data and the model spectra. The listed errors are arithmetic means for all spectra shown in Figure \ref{fig:atmos}. The average and relative rms errors (defined in Section \ref{sec:erroranalysis}) are 4\% and 6\% for altitudes greater than 12 km, at which cloud and H$_2$O variability become important. Our ATMOS spectral simulations did not include either clouds or multiple scattering, which have been shown to be important for modeling the reflected spectrum of the Earth \citep{robinson11}. Many of the 12 km spectra have increased extinction (lower transmittance levels) due to the presence of clouds truncating some of the paths. The ATMOS data at 12 km suggest that including a realistic treatment of clouds and multiple scattering may be necessary to adequately model transit transmission spectra of Earth-like exoplanets.

\begin{deluxetable}{ccc}
\tablecolumns{3}
\tablewidth{0pc}
\tablecaption{ATMOS Validation \label{tab:atmos}}
\tablehead{ \colhead{Height (km)} & \colhead{rms Error (\%)} & \colhead{rms Relative Error (\%)} }
\startdata
100 & 3.5 & 3.5 \\
75 & 3.1 & 3.1 \\
50 & 3.1 & 2.9 \\
40 & 4.1 & 2.8 \\
30 & 4.4 & 6.0 \\
25 & 4.4 & 9.3\\
20 & 5.8 & 12.5 \\
12 & 18.9 & $>$100\\
\hline
Average ($\geq$ 20km) & 4.0 & 5.7 \\
\enddata
\tablecomments{Absolute and relative rms errors for the model fits to the ATMOS data. The listed errors for each altitude are averaged over all the data sets available for that altitude. The data and model fit very well down to 12 km, at which point clouds and cloud variability increase dramatically. }
\end{deluxetable}

\subsubsection{Lunar Eclipse Validation}

We used the 2008 August lunar eclipse observations by \citet{palle09} to validate our model. The data were taken between 20:54 UT and 22:16 UT on 2008 August 16 at the observatory of El Roque de los Muchachos in La Palma Island using the LIRIS instrument on the William Herschel Telescope. The spectral resolution and range were $\sim$1000 and 0.9-1.5 $\mu$m for the observations we used for validation. During a lunar eclipse, the sunlight that reaches the Moon's surface traverses Earth's atmosphere on limb trajectories and is refracted, producing a spectrum similar to a transit transmission spectrum. The data are taken at different times throughout the transit with different angles between the Sun, Earth and Moon (the solar elevation angle) (\citet{garcia12}, Enric Pall{\'e}, private communication). The solar elevation angle (hereafter referred to as \textit{e}, in degrees) is the geocentric angle between the direction of incident sunlight and the direction of the moon from Earth's center. The amount of refraction necessary for light from the Sun to reach the lunar surface is largest at greatest eclipse (at which $e$ is at a minimum) and decreases as the Moon leaves greatest eclipse (as $e$ increases). The change in the necessary angle of refraction leads to variations in the lunar eclipse spectrum as the eclipse progresses, with more flux from progressively greater altitudes in Earth's atmosphere as $e$ increases.

\begin{figure}
\centering
\includegraphics[width=8cm]{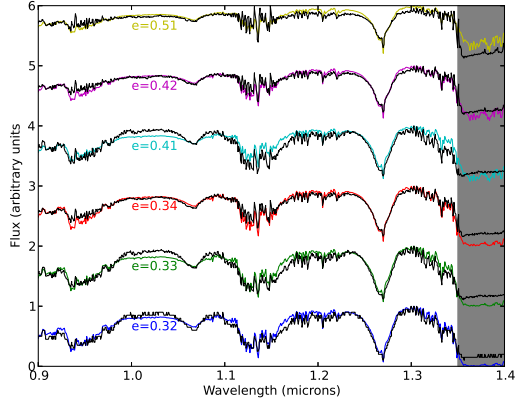}
\caption{Lunar eclipse spectra (black) and our model spectra (colors) at various geometries denoted by the solar elevation angle, $e$, in degrees. The shaded region denotes wavelengths with low photon counts at which the data flux levels should be considered only upper limits. Our model fits the data with errors less than the intrinsic discrepancies in the data due to atmospheric spatial and temporal variability and any observational errors. We consider the fits the lunar eclipse data to be a validation of the model, especially the refraction portion. \label{fig:palle-zj-all}}
\end{figure}

\begin{figure}
\centering
\includegraphics[width=8cm]{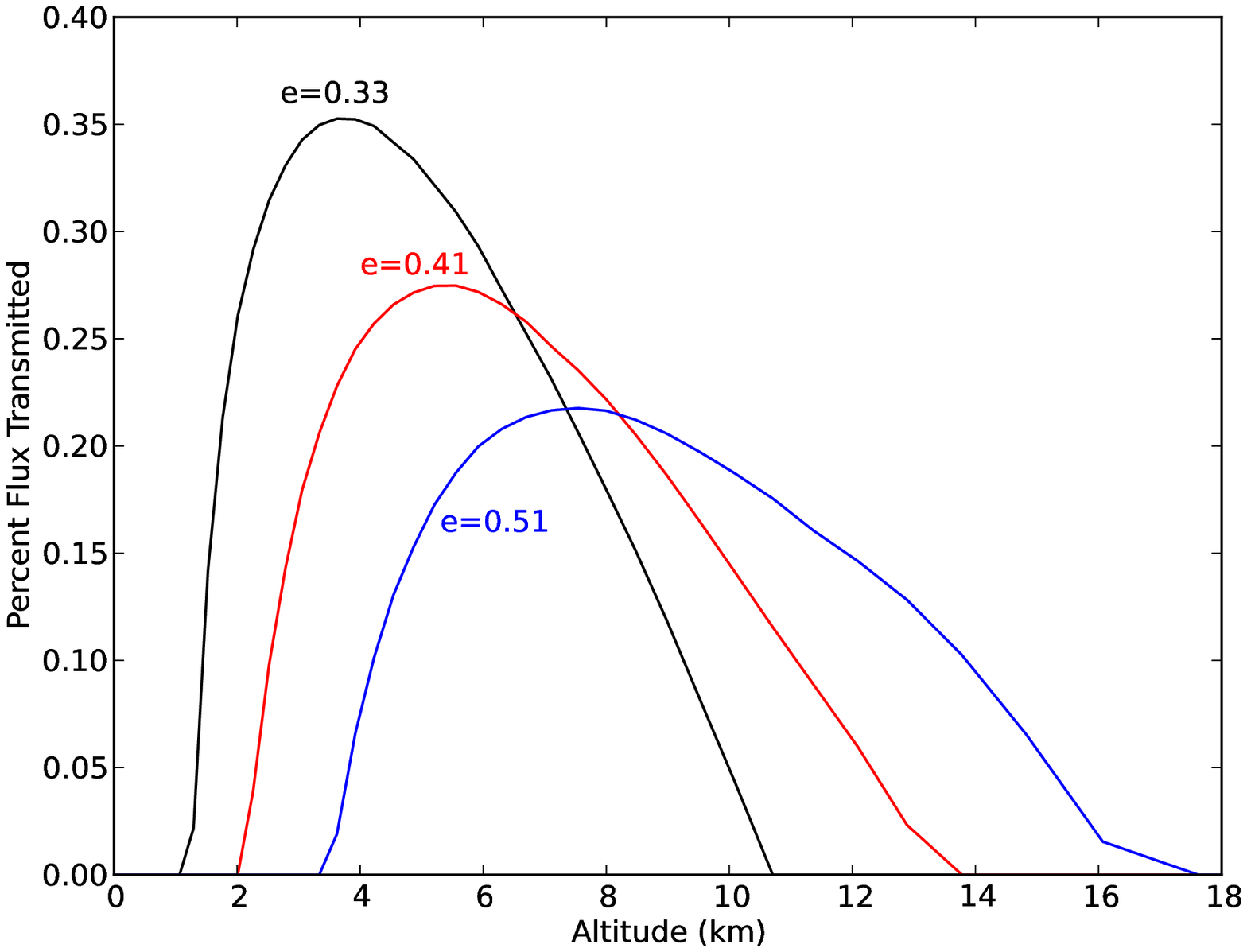}
\caption{Maximum transmitted flux at each altitude for the lunar eclipse spectra at different solar elevation angles ($e$). The results shown here are for the shortest wavelengths, but the wavelength dependence of the transmitted flux at each altitude is small. For the lunar eclipse spectra near greatest eclipse ($e$=0.33), only the lowest 11 km of the atmosphere were probed, and at the furthest point from greatest eclipse (e=0.51), altitudes between 3 and 18 km were probed. \label{fig:lunaralt} }
\end{figure}

Figure \ref{fig:palle-zj-all} shows a comparison of the \citet{palle09} lunar eclipse spectra and our model spectra. Table \ref{tab:pallevalid} shows the absolute and relative rms errors between the model and data. The data and model spectra typically agree to within $\sim$10\%. Figure \ref{fig:lunaralt} shows the altitudes that are probed at selected values of $e$ (the angle between the Sun, Earth, and Moon) for the lunar eclipse spectra. The lunar eclipse spectra typically probe altitudes between 2 and 18 km and only probe altitude between 1 and 11 km of the atmosphere near greatest eclipse.

\begin{deluxetable}{ccc}
\tablecaption{Lunar Eclipse Validation \label{tab:pallevalid} }
\tablewidth{0pc}
\tablecolumns{3}
\tablehead{
\colhead{$e$ (deg)} & \colhead{rms Error (\%)} & \colhead{rms Relative Error (\%)}
}
\startdata
0.32 & 5.8 & 10.5 \\
0.33 & 7.1 & 12.2 \\
0.34 & 5.1 & 9.2 \\
0.41 & 9.0 & 18.4 \\
0.42 & 5.0 & 7.1 \\
0.51 & 5.2 & 7.1 \\
\hline
Average & 6.2 & 10.8 \\
\enddata
\tablecomments{Absolute and relative errors for the model fits to the \citet{palle09} lunar eclipse observations. The intrinsic variability in the data corresponds to an absolute and relative error of 8.3\% and 14.4\%, respectively. Our model spectra generally agree to the data within these errors, and on average our errors are less than the intrinsic variability.}
\end{deluxetable}

We compared our errors in modeling the lunar eclipse spectra to the inherent variations in the data. The lunar eclipse spectra vary due to changes in the solar elevation angle, $e$. For geometries with similar $e$ values, the spectra should be nearly equivalent. Therefore, in order to quantify the inherent variance in the lunar eclipse data, we calculated the absolute and relative rms differences between spectra with close $e$ values. We compared the following cases: $e$=0.32$^{\circ}$, 0.33$^{\circ}$, 0.34$^{\circ}$ and $e$=0.41$^{\circ}$ and 0.42$^{\circ}$, and found that the average absolute and relative differences were 8.2\% and 14.2\%, respectively. These deviations are likely due to a combination of spatial and temporal variations in cloud coverage, aerosols, or gas mixing ratios and any observational errors. The lunar eclipse observations were taken at different times during the eclipse, such that different portions of the atmosphere were being probed as the Earth rotated and different cloud fields came into view. We used a single model atmosphere, so any spatial variations cannot be accounted for using our method. The average discrepancies of $\sim$8 and $\sim$14\% are slightly larger than our errors from our model validation of 7\% and 11\%, so we consider our fit to the lunar eclipse spectra to a be a validation of the model, especially the refraction portion.

\subsection{Maximum Tangent Pressure for Exoplanets} \label{sec:respressure}


Planetary atmospheres have index of refraction gradients which leads to the refraction, or deflection, of light as it passes through the planetary atmosphere. This refraction angle depends on the composition of the atmosphere, the pressure-temperature profile and the radius of the planet. For portions of the atmosphere with sufficiently large deflection angles, there will be no path originating at the host star that traverses the planetary atmosphere and can reach a distant observer, and these portions of the atmosphere will not be able to be probed by transit transmission spectroscopy. We define the maximum tangent pressure as the pressure at which 50\% of the flux is transmitted, which is roughly when the deflection angle is greater than the angular size of the host star as seen from the planet

\begin{figure}
\centering
\includegraphics[width=8cm]{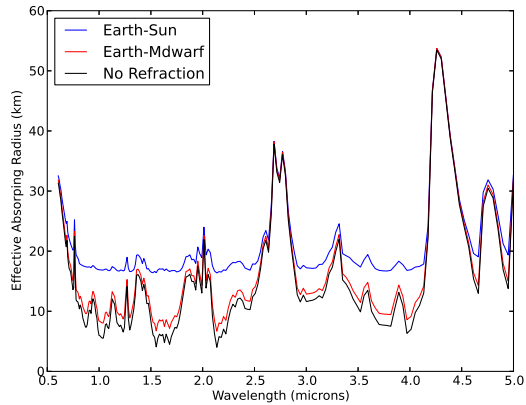}
\caption{Comparison of transit transmission spectra for an Earth analog orbiting a Sun-like star (blue) and an Earth analog orbiting an M dwarf (red), along with an Earth analog with no refractive effects included (black). The \textit{y} axis is the effective absorbing radius of the atmosphere (in km). As shown in Figure \ref{fig:maxtan}, the maximum tangent pressure that can be probed is $\sim$0.9 bars for the Earth-Sun analog and $\sim$0.3 bars for the Earth-M-dwarf system. This limit is set by refraction and leads to differences in the two spectra. Because it is possible to probe higher pressures for the M-dwarf case, the absorption features appear much stronger. Therefore, in general, planets orbiting M dwarfs should show stronger spectral signatures than planets orbiting more massive stars. \label{fig:earth-sun-mdwarf-spec}}
\end{figure}

\begin{figure}
\centering
\includegraphics[width=8cm]{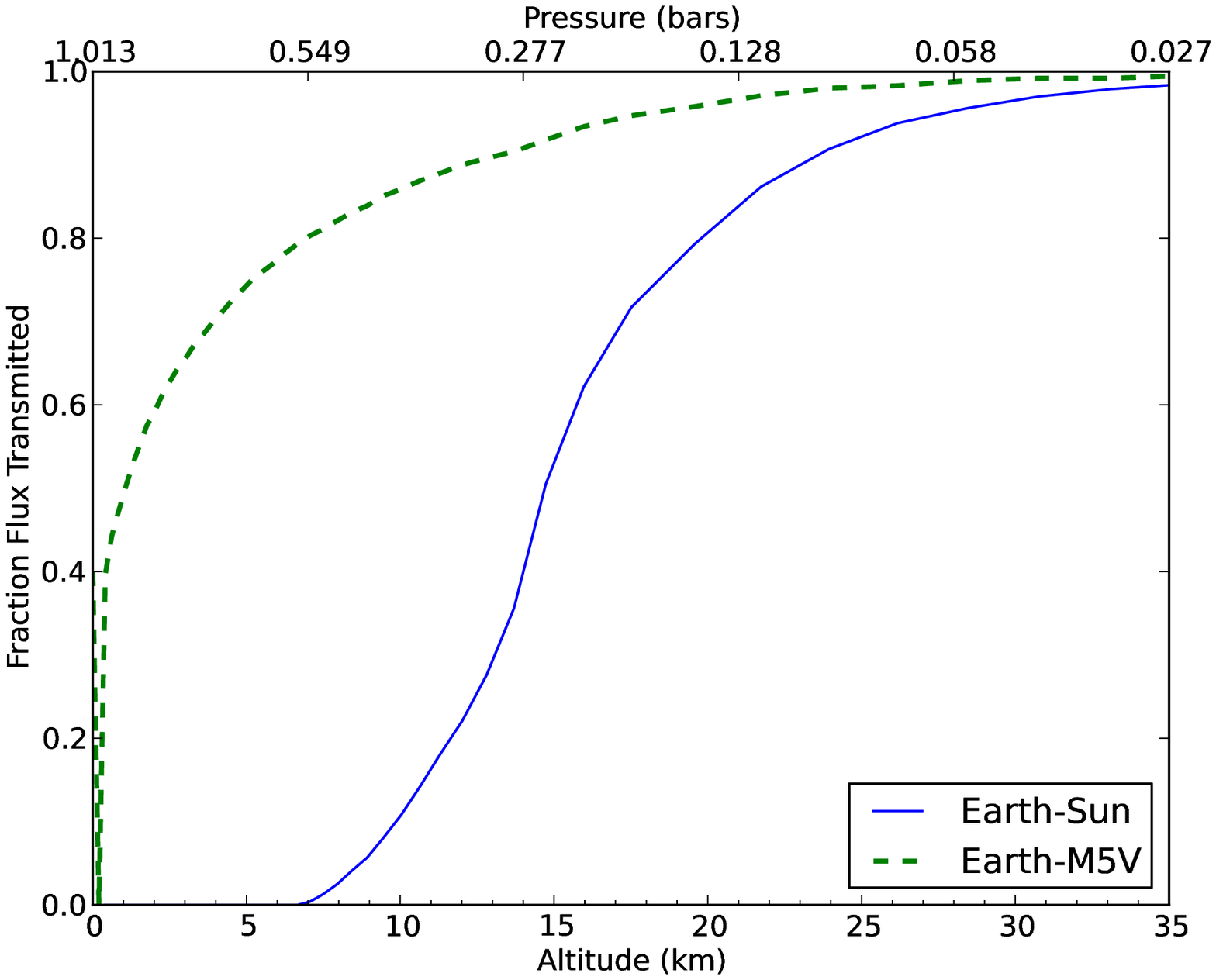}
\caption{Maximum amount of transmitted flux at each altitude for an Earth analog orbiting a Sun-like star and an Earth-analog orbiting an M5V star. The maximum flux is set by refraction and is independent of atmospheric opacity. For the Earth-Sun case, the lowest altitudes cannot be probed and the maximum tangent pressure (defined as the pressure at which 50\% of the flux is transmitted) is 0.3 bars. For the Earth-M5V case, almost all altitudes can be probed and the maximum tangent pressure is 0.9 bars. \label{fig:maxtan} }
\end{figure}

We explored the effect of the angular size of the star on the maximum tangent pressure cutoff for transit transmission spectra by running three test cases: an Earth analog orbiting a Sun-like star (angular size of .54$^{\circ}$), an Earth analog orbiting an M5V star (angular size of 2.2$^{\circ}$), and an Earth analog with no refraction (assuming a large enough angular size such that all pressures can be probed), with the resulting spectra shown in Figure \ref{fig:earth-sun-mdwarf-spec}. The maximum transmitted flux at each altitude, which is set by refraction, is shown in Figure \ref{fig:maxtan}. In the M-dwarf case, transit transmission spectroscopy can probe pressures as great as $\sim$0.9 bars (with a corresponding altitude of 1 km), while it can only probe up to $\sim$0.3 bars (14 km) for the Sun-like star case. The difference in maximum tangent pressure for the two cases produces absorption features that are much stronger for the M-dwarf case. 
 
\subsection{Temporal Variations Results}  \label{sec:sounding}

\begin{figure}
\centering
\includegraphics[width=8cm]{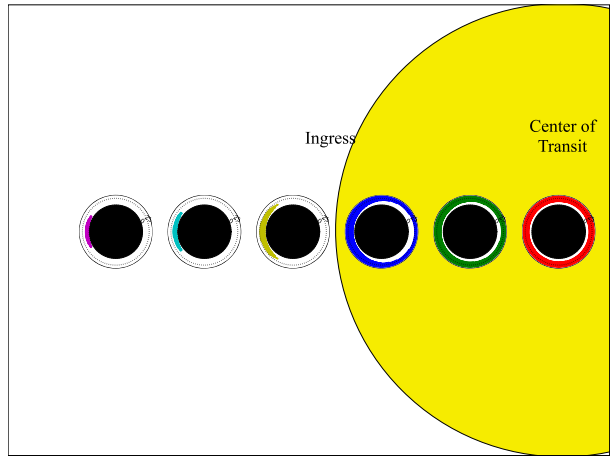}
\caption{Diagram showing which altitudes can be probed at different times during a transit for an Earth-Sun analog. The colored regions correspond to regions of the atmosphere where light is transmitted, and the white regions are portions of the atmosphere that are opaque to a distant observer. The colors correspond to the spectra colors in Figures \ref{fig:vary-avg} and \ref{fig:vary-diff}. Prior to ingress, only low altitudes in the atmosphere can be probed. As the planet moves from ingress to center of transit, more flux can be transmitted through progressively higher pressures (or lower altitudes) until the maximum tangent pressure is reached. \label{fig:vary-diagram} }
\end{figure}

\begin{figure}
\centering
\includegraphics[width=8cm]{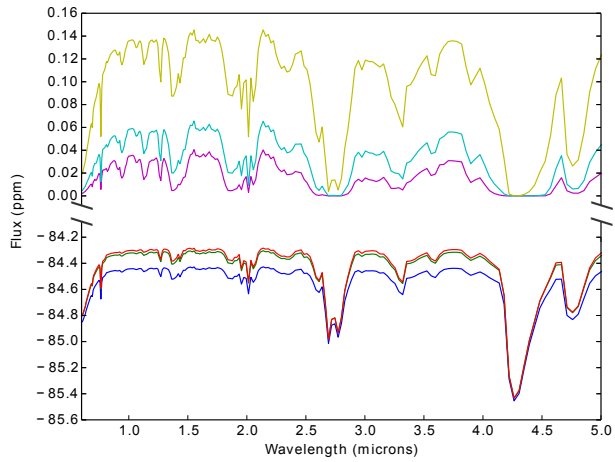}
\caption{Time-averaged spectra for the out-of-transit (top) and in-transit (bottom) time periods illustrated in Figure \ref{fig:vary-diagram} during a transit of an Earth-Sun analog. The spectra are shown in terms of ppm flux differences. The largest variability (besides ingress) is seen between ingress and one third of the way to center-of-transit. The upper scale has been expanded to better display the out-of-transit spectra. \label{fig:vary-avg}}
\end{figure}

\begin{figure*}
\centering
\includegraphics[width=16.5cm]{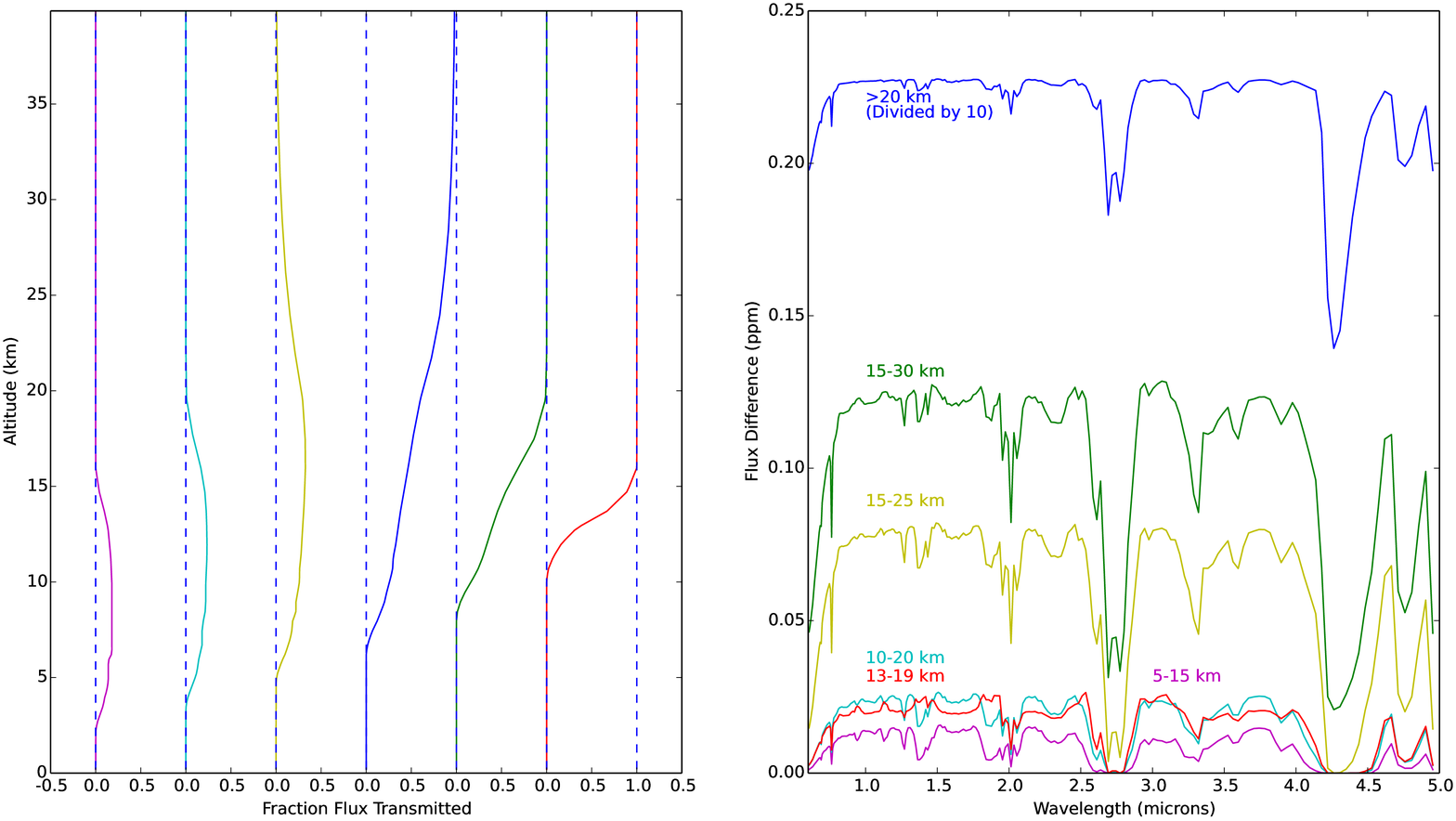}
\caption{Altitude-dependent transmitted flux and spectra for an Earth-like planet orbiting a Sun-like star. Left: altitude-dependent transmitted flux from pre-transit to center of transit as deviations from the dotted line. Right: The purple spectrum is plotted as would be observed, and all other spectra show the spectrum of that stage with the previous stage's spectrum subtracted from it, e.g., the cyan spectrum shows the difference between the spectrum in the cyan and purple stages. The six stages correspond to the stages and colors shown in Figure \ref{fig:vary-diagram}. Taking the difference between the spectra at each stage allows us to isolate a specific vertical region of the atmosphere, permitting vertical sounding of the atmosphere. \label{fig:vary-diff}}
\end{figure*}

Figures \ref{fig:vary-diagram}-\ref{fig:vary-diff} show how differences in the portions of the exoplanet atmosphere that are backlit (meaning light is transmitted to a distant observer) as a transit progresses can lead to temporal variations in a transit transmission spectrum of an exoplanet. These figures show the spectra for an Earth-analog planet orbiting a Sun-like star from half a transit length prior to ingress to center of transit. The colors in each figure correspond to a different stage of the transit. During the earliest stage (purple), only a small portion on the trailing (left) portion of the limb is backlit, as shown in Figure \ref{fig:vary-diagram}. The planet is not in transit yet, so most of the atmosphere is not backlit and the portions that are backlit are at altitudes ($\sim$2-15 km) at which the deflection angle is large enough that light from the far (right side) limb of the star is deflected into the beam to a distant observer. Most of the atmosphere is opaque, and therefore the flux transmitted through the atmosphere is small, as shown in Figure \ref{fig:vary-avg}. The altitudes that are probed in this stage (and all other stages) are shown in the left-hand side of Figure \ref{fig:vary-diff}. In the next two stages of the transit (cyan and yellow), the angle of deflection required for light to reach a distant observer is smaller than in the first (purple) stage, so the portions of the atmosphere that are backlit are at progressively greater altitudes ($\sim$5-17 km and 5-30 km, respectively). 

After ingress (blue), all altitudes above $\sim$7 km are at least partially backlit and more flux is transmitted through the atmosphere. As shown in the left-hand side of Figure \ref{fig:vary-diff}, the upper atmosphere is entirely backlit, but portions of the lower atmosphere are deflecting light out of the beam to a distant observer. As the planet reaches center of transit (green, then red), more flux can be transmitted at altitudes between 15 and 25 km, and slightly less flux is transmitted at altitudes between 10 and 15 km. The overall absorbing radius decreases because the net result of these flux changes is that more flux is transmitted through the atmosphere. We note that at center of transit, no pressures greater than the maximum tangent pressure can be probed. However, prior to ingress, it is possible to probe pressures greater than the maximum tangent pressure, and, in fact all the flux transmitted during the first stage (purple) is transmitted for paths with tangent pressures greater than the maximum tangent pressure.

The differences in the spectra as the planet moves through the stages of its transit can be used to obtain vertical sounding, or altitude-dependent spectra of the exoplanet. The right-hand side of Figure \ref{fig:vary-diff} shows the successive differences in the spectra at each stage of the transit, except for the first (purple) spectrum which is plotted as it would be observed, and can be considered the difference between the spectrum in the first stage and the spectrum well before the planet approaches ingress. We take the difference of the spectra to isolate a specific vertical region of the atmosphere. For example, the spectrum shown in yellow for the third stage is the difference between the spectrum for the third (yellow) and second (cyan) stage of the transit and corresponds to flux transmitted at tangent altitudes between 15 and 25 km. By looking at the differences in the spectra between each stage of the transit, it could be possible to retrieve altitude-dependent mixing ratios for gases in the exoplanet atmosphere.

\begin{figure*}
\centering
\includegraphics[width=16.5cm]{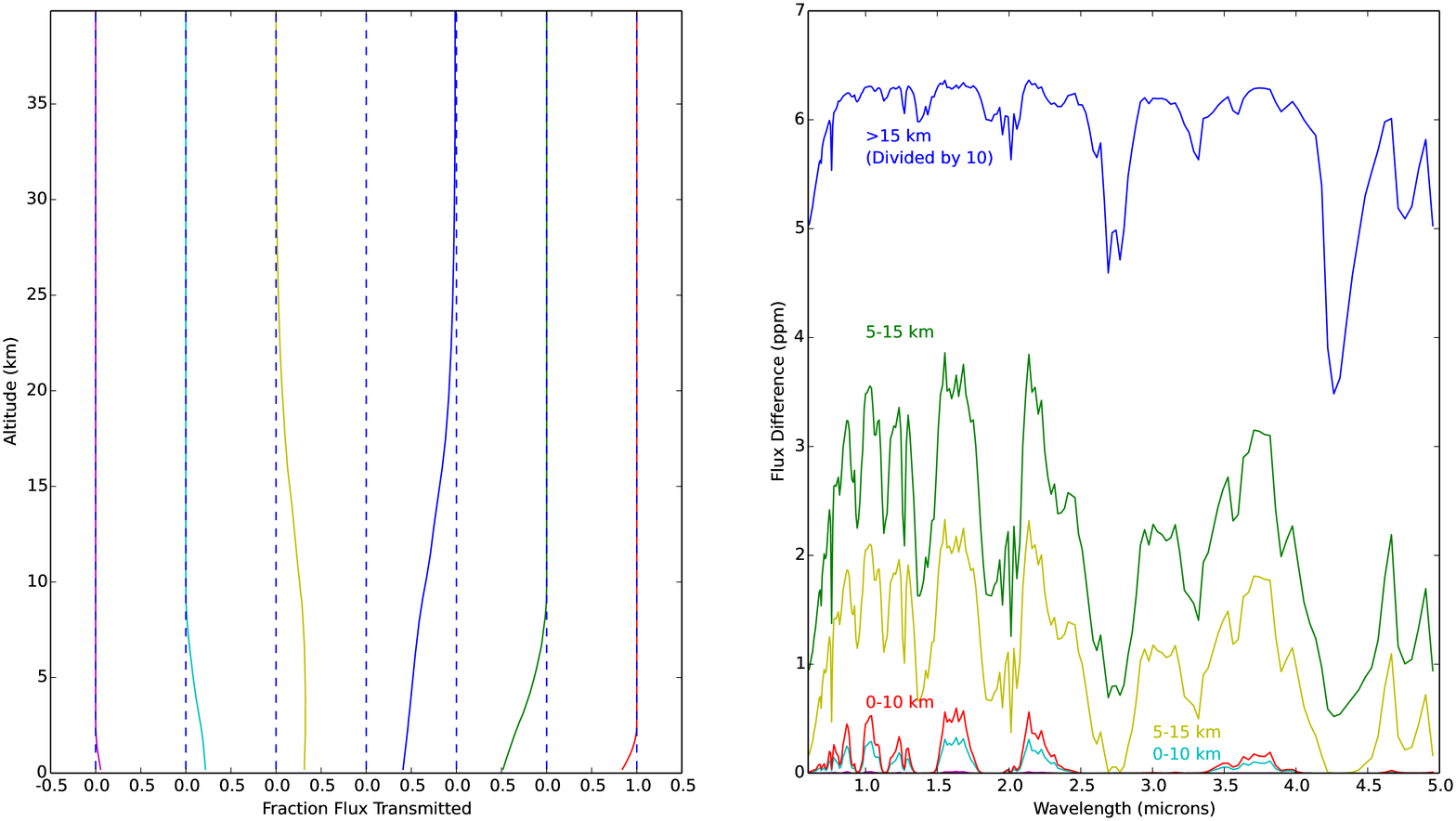}
\caption{Same as Figure \ref{fig:vary-diff}, but for an Earth analog orbiting an M dwarf. The differences in flux occur at higher pressures than for the Earth-Sun case because transit transmission can probe higher pressures for the Earth-M-dwarf case. \label{fig:vary-diff-mdwarf}}
\end{figure*}

Figure \ref{fig:vary-diff-mdwarf} shows the temporal variations for an Earth analog orbiting an M-dwarf. The variations correlate to differences at lower altitudes but have a greater overall signal. The absorption features are much deeper than for the Earth-Sun case because the flux differences correspond to lower altitudes or greater pressures being probed.

\subsection{Detectability of Spectral Features} \label{sec:detect}

\begin{figure}
\centering
\includegraphics[width=8cm]{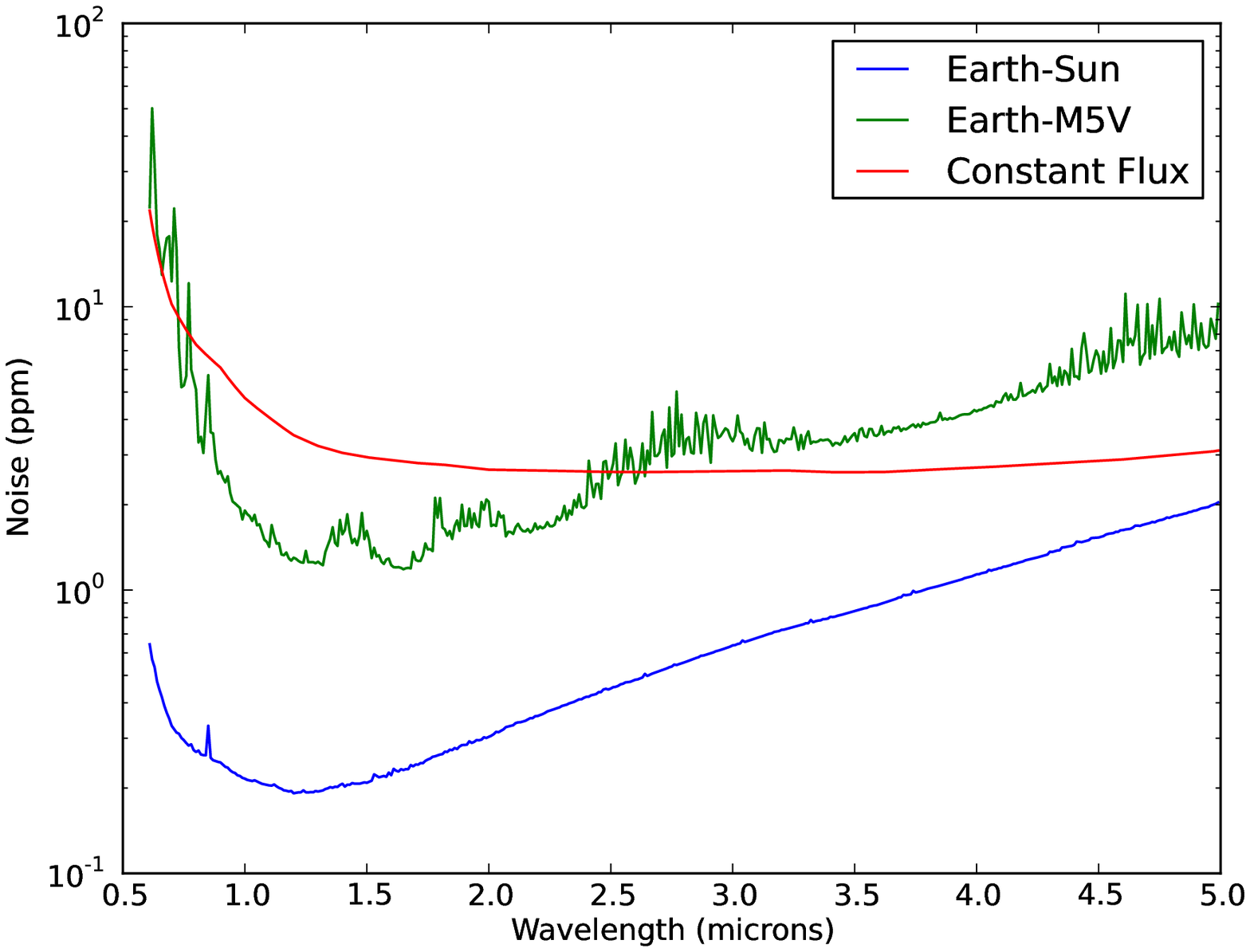}
\caption{Noise estimates (in ppm) for an Earth-Sun analog, an Earth-M5V analog, and a flat continuum source of 5 $\times$10$^-11$W m$^{-2}$ $\mu$m$^{-1}$. The exposure times are 234,000 s for the Earth-Sun case and 1,000,000 s for the others. \label{fig:jwstnoise}}
\end{figure}

We calculated the S/Ns of spectral absorption features that could be achieved with the \textit{JWST} NIRSPEC instrument, assuming all transits in the five-year lifetime of \textit{JWST} could be observed. Figure \ref{fig:jwstnoise} shows the noise levels for the Earth-Sun analog, Earth-M5V analog and a flat continuum source. The noise levels vary with wavelength because the stellar fluxes are wavelength-dependent and also because the NIRSPEC sensitivity is wavelength-dependent. Table \ref{tab:jwstnoise} shows the S/N and signal level for a number of spectral features for four test cases: the Earth-Sun analog and Earth-M5V analog, each with and without refraction. Many of the spectral features identified in the transit transmission spectrum of an Earth analog around an M5V star could be detected by \textit{JWST} at the 3$\sigma$ level or greater, assuming all transits in the 5 year mission lifetime are coadded. Table \ref{tab:jwstnoise} shows that there are H$_2$O and CO$_2$ features with S/Ns greater than seven. Furthermore, O$_2$ and CH$_4$, which together are considered a very strong biosignature \citep{lovelock65}, are detectable with S/Ns of $\sim$3 for the 1.27 $\mu$m O$_2$ band and the 2.3 and 3.3 $\mu$m CH$_4$ bands. 

Our results generally agree with previous detectability calculations. \citet{deming09} find S/Ns for the 4.3 $\mu$m CO$_2$ band and H$_2$O bands between 1.7 and 3.0 $\mu$m were near 50 for habitable super-Earths at a distance of 10 pc. These S/N values are a few times higher than what we have found, but our calculations were done for an Earth-like atmosphere, while theirs were done for super-Earths with different compositions, some of which had greater amounts of H$_2$O and CO$_2$ than the Earth. \citet{kalt09} find much higher S/N values for their Earth-analog calculations, but we note that they do not include accurate noise estimates for \textit{JWST} and do not include a factor of $\sqrt{2}$ in their calculations for the noise added when differencing in-transit and out-of-transit observations. \citet{belu11} find an S/N of between 5 and 10 for the 4.3 $\mu$m CO$_2$ band for a super-Earth, which is consistent with what we find for a true Earth analog. 

\begin{deluxetable}{llllllllll}
\tablecaption{S/Ns for Absorption Features \label{tab:jwstnoise}}
\tablecolumns{10}
\tablewidth{0pc}
\tablehead{
\multicolumn{2}{c}{} & \multicolumn{4}{c}{With Refraction} & \multicolumn{4}{c}{No Refraction} \\ \hline
\multicolumn{2}{c}{Absorption Feature} & \multicolumn{2}{c}{ Earth-Sun} & \multicolumn{2}{c}{Earth-M5V} & \multicolumn{2}{c}{Earth-Sun} & \multicolumn{2}{c}{Earth-M5V}\\ \hline
\colhead{Molecule} & \colhead{$\lambda$ ($\mu$m)} & \colhead{S/N} & \colhead{S} & \colhead{S/N} & \colhead{S} & \colhead{S/N} & \colhead{S} & \colhead{S/N} & \colhead{S} 
}
\startdata
O2 & 0.69 & 0.1 & 0.0 & 0.0 & 1.1 & 0.1 & 0.0 & 0.1 & 1.2 \\ 
O2 & 0.76 & 0.4 & 0.1 & 0.4 & 6.2 & 0.7 & 0.3 & 0.5 & 6.7\\ 
O2-O2 & 1.06 & 0.0 & 0.0 & 0.9 & 2.0 & 0.4 & 0.1 & 1.2 & 2.7 \\
O2 & 1.27 & 0.3 & 0.1 & 2.9 & 5.1 & 0.9 & 0.2 & 3.3 & 5.9 \\
O3 & 4.70 & 0.2 & 0.6 & 1.6 & 18.5 & 0.3 & 0.8 & 1.7 & 19.3 \\
CO2 & 2.00 & 0.7 & 0.3 & 7.5 & 19.7 & 2.1 & 0.9 & 8.7 & 22.8\\
CO2 & 2.70 & 1.5 & 1.1 & 7.5 & 33.7 & 1.9 & 1.4 & 7.8 & 34.9\\
CO2 & 4.30 & 1.0 & 1.9 & 7.4 & 58.4 & 1.3 & 2.4 & 7.8 & 61.0\\
CH4 & 2.30 & 0.2 & 0.1 & 4.3 & 12.0 & 1.0 & 0.6 & 5.1 & 14.6\\
CH4 & 3.30 & 0.3 & 0.3 & 3.5 & 17.2 & 0.7 & 0.8 & 3.8 & 18.9\\
H2O & 0.90 & 0.1 & 0.0 & 2.0 & 6.2 & 1.0 & 0.3 & 2.5 & 7.9\\
H2O & 1.10 & 0.1 & 0.0 & 2.6 & 5.2 & 0.9 & 0.3 & 3.1 & 6.3\\
H2O & 1.40 & 0.6 & 0.2 & 6.6 & 15.6 & 2.5 & 0.7 & 7.9 & 18.3\\
H2O & 1.90 & 0.4 & 0.2 & 6.6 & 16.3 & 2.0 & 0.8 & 8.0 & 19.6\\
H2O & 2.70 & 0.4 & 0.3 & 2.9 & 12.3 & 0.8 & 0.5 & 3.1 & 13.2\\
\enddata
\tablecomments{Signal levels (in ppm) and signal to noise ratio (S/N) for molecular absorption features for four different test cases, for the Earth-Sun analog and the Earth-M5V analog, each with and without refraction included. The calculations were done assuming that all possible transits are observed in \textit{JWST}'s 5 year mission lifetime.}
\end{deluxetable}

\begin{figure}
\centering
\includegraphics[width=8cm]{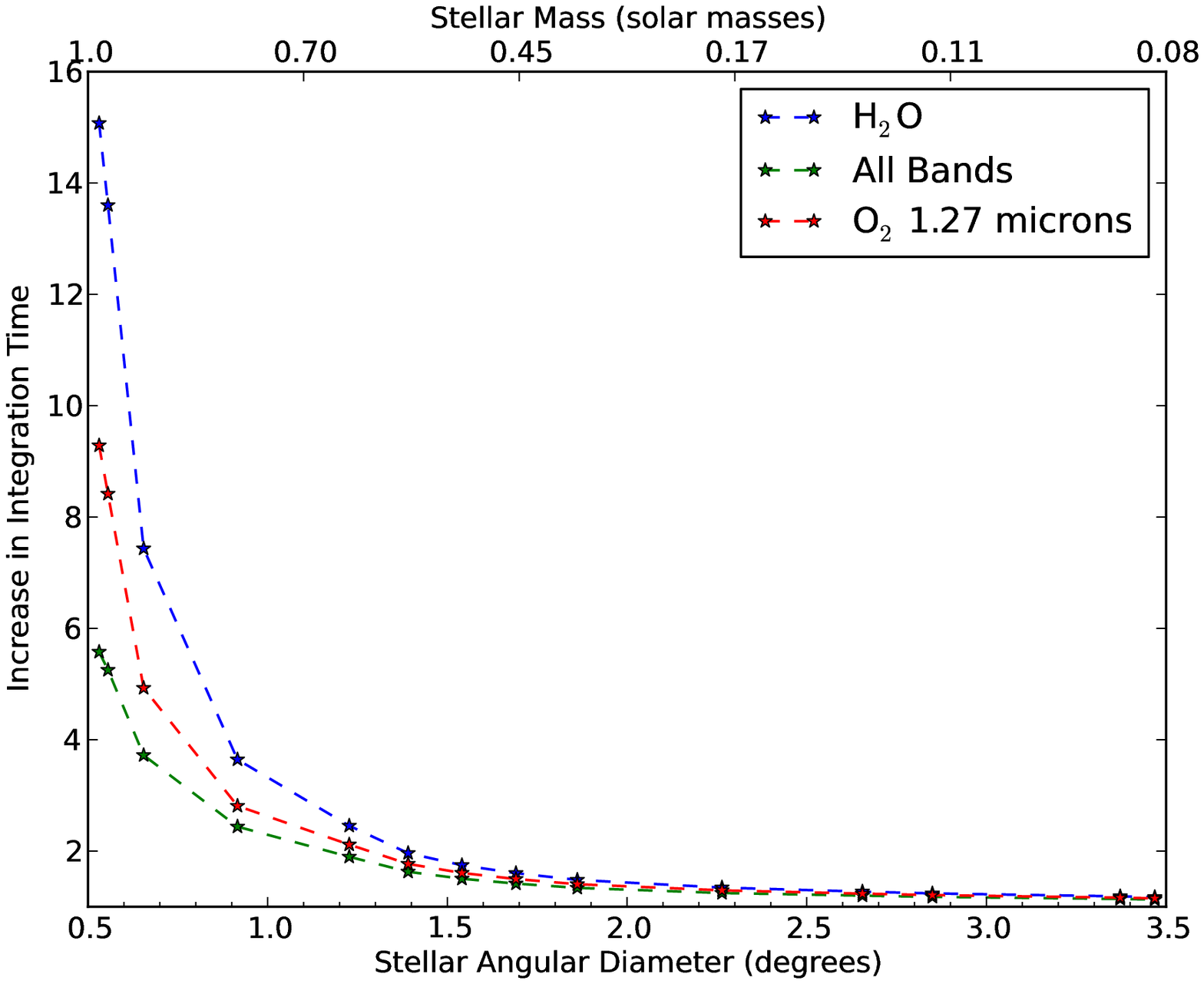}
\caption{Fractional increase in integration time required to achieve the same S/N for absorption features in a refracted spectrum compared with a non-refracted spectrum versus the angular diameter of the star from the Earth analog's perspective. The stellar masses given on the top \textit{x} axis are the mass of the star for a given angular diameter assuming the Earth analog is orbiting at a flux-equivalent distance, or the distance where the top of atmosphere incident flux is 1373 W m$^{-2}$. Molecules that are concentrated at the lowest altitudes, such as H$_2$O and dimer molecules, are the most affected by the cutoff set by refraction. \label{fig:inttime}}
\end{figure}

Refraction limits the detectability of spectral absorption features for all cases examined here, with a greater effect for Earth-like planets orbiting Sun-like stars. For the Earth-M5V case, refraction leads to a decrease in the predicted S/N by an average of $\sim$10\% for all spectral features and $\sim$15\% for H$_2$O features. However, for the Earth-Sun case, refraction leads to an average decrease in the S/N by $\sim$60\% for all features and $\sim$75\% for H$_2$O features. We have further quantified this effect in Figure \ref{fig:inttime}, which shows the fractional increase in integration time required to obtain the same S/N for spectral features in a refracted spectrum compared to a non-refracted spectrum as a function of the angular diameter of the star from the planet's perspective. We find that molecules that are concentrated at the lowest altitudes, such as H$_2$O and O$_2$ dimer molecules, would require the greatest increase in integration time to be detected for the refracted model spectra. Evenly mixed molecules (like CO$_2$ and O$_2$) and molecules concentrated in the stratosphere (O$_3$) typically require lower increases in integration time. For all molecules, planets orbiting M-dwarf stars require the smallest increases in integration time. In contrast, a factor of 16 increase in integration time would be required to detect multiple H$_2$O absorption features in the spectrum of an Earth-like planet orbiting a Sun-like star.

\textit{JWST} will not be able to detect obtain vertical sounding for an Earth-like planet. The S/Ns for even the strongest features in any of the temporally varying spectra are no greater than $\sim$1.7 for the Earth-M5V case and far lower for the Earth-Sun case. However, we find that by integrating over all wavelengths, it may be possible to detect broadband temporal variations due to refraction with S/Ns of 4.6 (out of transit) and 8.1 (in transit) for the Earth-M5V case and 1.5 (out of transit) and 2.3 (in transit) for the Earth-Sun case. The out-of-transit broadband refracted light signal would be an increase in flux just prior to ingress and subsequent to egress in the transit light curve, while the in-transit refracted light signal would be an increase in flux from ingress to center of transit, then a decrease from center of transit to egress. While detecting broadband variations would not reveal anything about the vertical mixing ratio profiles of atmospheric gases, it could potentially be used to obtain information on the vertical structure of the atmosphere and would be a confirmation of the model predictions.

\section{Discussion}



As described in Section \ref{sec:respressure}, the maximum tangent pressure that can be probed is greater for larger apparent angular stellar sizes and is greater for less refractive atmospheres. For the exoplanets whose transit transmission spectra have been recorded to date, the maximum tangent pressures have been large (typically $\geq10$ bars) due to the small planet-star distances (and thus large apparent stellar diameters). At these pressures, all atmospheres should be opaque due to collision-induced absorption or pressure broadening, suggesting that all optically thin paths will be largely unaffected by refraction. Therefore, the effect of refraction is small for most observations of very close-in planets, such as hot Jupiters of hot Neptunes.


The maximum tangent pressure can determine how observable a biosignature could be for an exoplanet. As shown in Figure \ref{fig:earth-sun-mdwarf-spec}, Table \ref{tab:jwstnoise} and Figure \ref{fig:inttime}, absorption features for a habitable planet orbiting an M dwarf are much stronger than for a habitable planet orbiting a Sun-like star because most of the atmosphere can be probed for an Earth analog orbiting an M dwarf, while only pressures less than 0.3 bars can be probed for an Earth analog orbiting a Sun-like star. This also has implications for potentially habitable planets orbiting white dwarfs. Planets in the habitable zones of white dwarfs could show very strong transit transmission signals \citep{agol11}. Additionally, white dwarfs have similar temperatures to the Sun, and the angular diameter of the white dwarf for a planet in the habitable zone is roughly the angular diameter of the Sun from the Earth's perspective. \citet{loeb13} find that five hours of \textit{JWST} integration time are required to detect the O$_2$ \textit{A} band on an Earth-like planet orbiting in the habitable zone of a white dwarf, but because of refraction we calculate that the integration time will increase by as much as a factor of six for detecting the O$_2$ \textit{A} band for Earth analogs orbiting white dwarfs.



We demonstrate how temporal variations in a transit transmission spectrum can be used to obtain vertical sounding in exoplanet atmospheres (see Figures \ref{fig:vary-diagram}-\ref{fig:vary-diff}). Vertical sounding in an exoplanet atmosphere can provide a wealth of information relevant to atmospheric structure. Vertical profiles of gas mixing ratios can reveal whether or not a gas is evenly mixed in an atmosphere. Unevenly mixed gases, such as CH$_4$, O$_3$, and H$_2$O in Earth's atmosphere, can provide information about surface fluxes, photochemistry, UV shielding, and the presence of a cold trap, or a layer in the atmosphere at which water condenses and forms clouds. \textit{JWST} will likely not be able to use the techniques described here to obtain vertical sounding of an Earth-like exoplanet at 10 pc, as the S/Ns for these features are all below two (see Section \ref{sec:detect}). Therefore, these types of observations will not be possible until noise levels of a few times lower than those shown in Figure \ref{fig:jwstnoise} can be achieved, which would require a larger ground or space-based observatory. 

A detection of temporal variations would be most robust when detected over multiple transits and if stellar variability could be ruled out by looking for a wavelength-dependent, repeatable signal. Starspots and other forms of stellar variability could produce temporal variations in a transit transmission spectrum. As a transiting exoplanet passes over a starspot, the total amount of flux blocked from the planet decreases because starspots tend to be cooler than the stellar photosphere. This leads to an increase in total flux. Similarly, refraction leads to a an increase in flux, or decrease in absorption, as the planet nears center of transit. For the out-of-transit variations, a starspot moving across the stellar disk could potentially lead to a an increase in flux just prior to ingress or just subsequent to egress, which would mimic the signal predicted from our refraction model. For either the in-transit or out-of-transit variations, the wavelength dependence of the variations could be used to distinguish between refraction and stellar variability because variations due to refraction will exhibit spectral absorption features from the planetary atmosphere (see Figures \ref{fig:vary-avg} and \ref{fig:vary-diff}(B)). In comparison, starspot spectra are similar to blackbodies with effective temperatures lower than that of the rest of the stellar photosphere. Therefore, the wavelength dependence of the temporal variations could be used to distinguish between stellar variability and a signal due to refraction, potentially even with observations in two broadband filters if necessary.

Refractive effects will also be more repeatable than those of stellar variability. Starspots have limited lifetimes and will rotate in and out of view. The rotation periods of planet-hosting stars are on the order of days, and longer for M dwarfs, \citep{mcquilan13, nielsen13}. For the cases considered here, the transit duration is on the order of hours and the transit periods are typically tens of days for planets orbiting M dwarfs and hundreds of days for planets orbiting earlier-type stars. Because the transit duration is generally much shorter than the stellar rotation period, the starspot pattern on a star will be relatively static over the course of a transit and will unlikely be able to produce temporal variations similar to out-of-transit variations due to refraction. Furthermore, the starspot pattern on a star will be different for each transit, making it improbable that temporal variations due to stellar variability would be seen over multiple transits. On the other hand, variations due to refraction should be constant between transits, meaning the repeatability of the variations could aid in distinguishing between stellar variability and refractive effects.


\section{Conclusions}

We have presented a new model for transit transmission spectra that includes refraction. This model has been validated against ATMOS data and lunar eclipse spectra. We used the model to quantify how refraction limits the atmospheric pressures levels that can be probed and find that refraction sets a maximum tangent pressure of 0.9 bars for an Earth analog orbiting a M5V star and 0.3 bars for an Earth analog orbiting a Sun-like star. Furthermore, because of refraction, the S/N of absorption features in transit transmission can decrease by $\sim$10\% for the Earth-M5V case and $\sim$60\% for the Earth-Sun case. Additionally, detecting biosignature molecules such as O$_2$ for the Earth-Sun case will require as much as 10 times greater integration time compared to predictions from models run without refraction. Potentially habitable planets orbiting white dwarfs will have a similar pressure cutoff as for an Earth analog orbiting a Sun-like star. For planets in the habitable zone, refraction will have the smallest effect for detecting absorption features on planets orbiting M dwarfs. We find that despite the S/N decrease due to refraction, CO$_2$, H$_2$O, CH$_4$ and possibly O$_2$ are detectable at the 3$\sigma$ level for an Earth analog orbiting an M5V star 10 pc away in the habitable zone if every possible transit is observed over the five-year lifetime of \textit{JWST}.

We calculated the temporal variations in a transit transmission spectrum due to changes in the regions of the planetary atmosphere that are backlit at different stages of a transit. These variations could potentially allow us to obtain altitude-dependent spectra of an exoplanet, which can be used to retrieve vertical profiles for gas mixing ratios in exoplanet atmospheres. \textit{JWST} will most likely not be able to acquire the required S/N to obtain vertical profiles of absorbing gases for Earth-like planets at 10 pc. Therefore, obtaining vertical sounding for Earth-like exoplanets will likely require future ground and space-based observatories that can achieve lower noise levels than \textit{JWST}.

\section*{Acknowledgments}

We thank Bill Irion and Michael Gunson for helpful discussions regarding the ATMOS mission and data. We thank Enric Pall{\'e} for providing us with the lunar eclipse data.

This work was performed by the NASA Astrobiology Institute's Virtual Planetary Laboratory, supported by the National Aeronautics and Space Administration through the NASA Astrobiology Institute under Cooperative Agreement solicitation NNH05ZDA001C.  This work has also been supported by a generous fellowship from the ARCS Seattle chapter and funding from the Astrobiology program at the University of Washington under an NSF IGERT award.

Some of the work described here was conducted at the Jet Propulsion Laboratory, California Institute of Technology, under contract with NASA.

This research has made use of NASA's Astrophysics Data System.


\begin{thebibliography}

\bibitem[Agol(2011)]{agol11} Agol, E.\ 2011, \apjl, 731, L31 






\bibitem[Belu et 
al.(2011)]{belu11} Belu, A.~R., Selsis, F., Morales, J.-C., et al.\ 2011, \aap, 525, A83 


\bibitem[Bohren 
\& Huffman(1983)]{bohren83} Bohren, C.~F., \& Huffman, D.~R.\ 1983, New York: Wiley, 1983, 


\bibitem[Bourassa et al.(2010)]{bourassa10} Bourassa, A.~E., 
Degenstein, D.~A., Elash, B.~J., 
\& Llewellyn, E.~J.\ 2010, Journal of Geophysical Research (Atmospheres), 115, D00L03


\bibitem[Brown(2001)]{brown01} Brown, T.~M.\ 2001, \apj, 553, 
1006 


\bibitem[Charbonneau et al.(2002)]{char02} Charbonneau, D., 
Brown, T.~M., Noyes, R.~W., \& Gilliland, R.~L.\ 2002, \apj, 568, 377 


\bibitem[Claret(2000)]{claret00} Claret, A.\ 2000, \aap, 363, 1081 


\bibitem[Clough et al.(1992)]{clough92} Clough, S.~A., Iacono, 
M.~J., \& Moncet, J.-L.\ 1992, \jgr, 97, 15761 


\bibitem[Crisp(1997)]{crisp97} Crisp, D.\ 1997, \grl, 24, 571 


\bibitem[Deming et al.(2009)]{deming09} Deming, D., Seager, S., 
Winn, J., et al.\ 2009, \pasp, 121, 952 




\bibitem[Elliot 
\& Olkin(1996)]{elliot96} Elliot, J.~L., \& Olkin, C.~B.\ 1996, Annual Review of Earth and Planetary Sciences, 24, 89 

\bibitem[Fedorova et al.(2009)]{fedorova09} Fedorova, A.~A., 
Korablev, O.~I., Bertaux, J.-L., et al.\ 2009, \icarus, 200, 96 


\bibitem[Garc{\'{\i}}a Mu{\~n}oz et al.(2012)]{garcia12} 
Garc{\'{\i}}a Mu{\~n}oz, A., Zapatero Osorio, M.~R., Barrena, R., et al.\ 
2012, \apj, 755, 103 


\bibitem[Greenblatt et al.(1990)]{greenblatt90} Greenblatt, G.~D., 
Orlando, J.~J., Burkholder, J.~B., 
\& Ravishankara, A.~R.\ 1990, \jgr, 95, 18577 


\bibitem[Gunson et al.(1990)]{gunson90} Gunson, M.~R., Farmer, 
C.~B., Norton, R.~H., Zander, R., \& Rinsland, C.~P.\ 1990, \jgr, 95, 13867 


\bibitem[Hauschildt et al.(1999)]{hauschildt99} Hauschildt, P.~H., 
Allard, F., \& Baron, E.\ 1999, \apj, 512, 377 




\bibitem[Hubbard et al.(2001)]{hubbard01} Hubbard, W.~B., 
Fortney, J.~J., Lunine, J.~I., et al.\ 2001, \apj, 560, 413 


\bibitem[Hui 
\& Seager(2002)]{hui02} Hui, L., \& Seager, S.\ 2002, \apj, 572, 540 


\bibitem[Irion et al.(2002)]{irion02} Irion, F.~W., Gunson, 
M.~R., Toon, G.~C., et al.\ 2002, \ao, 41, 6968 


\bibitem[K{\"o}hler et al.(2005)]{kohler05} K{\"o}hler, J., 
Melf, M., Posselt, W., Holota, W., 
\& te Plate, M.\ 2005, \procspie, 5962, 563 


\bibitem[Kaltenegger 
\& Traub(2009)]{kalt09} Kaltenegger, L., \& Traub, W.~A.\ 2009, \apj, 698, 519 

\bibitem[Knutson et al.(2014)]{knutson14} Knutson, H.~A., 
Benneke, B., Deming, D., \& Homeier, D.\ 2014, \nat, 505, 66 

\bibitem[Kreidberg et al.(2014)]{kreidberg14} Kreidberg, L., Bean, 
J.~L., D{\'e}sert, J.-M., et al.\ 2014, \nat, 505, 69 

\bibitem[Kurucz(1979)]{kurucz79} Kurucz, R.~L.\ 1979, \apjs, 40, 
1 


\bibitem[Lafferty et al.(1996)]{lafferty96} Lafferty, W.~J., 
Solodov, A.~M., Weber, A., Olson, W.~B., 
\& Hartmann, J.-M.\ 1996, \ao, 35, 5911 


\bibitem[Loeb 
\& Maoz(2013)]{loeb13} Loeb, A., \& Maoz, D.\ 2013, \mnras, 432, L11 


\bibitem[Lovelock(1965)]{lovelock65} Lovelock, J.~E.\ 1965, \nat, 
207, 568 


\bibitem[Mat{\'e} et al.(1999)]{mate99} Mat{\'e}, B., Lugez, 
C., Fraser, G.~T., \& Lafferty, W.~J.\ 1999, \jgr, 104, 30585 

\bibitem[McClatchey et al.(1972)]{mcclatchey72} McClatchey, R. A., Fenn, R. W., Selby, J. E. A., Volz, F. E., Garing, J. S. 1972. Optical Properties of the Atmosphere (Third Edition). Air Force Cambridge Research Labs.

\bibitem[McQuillan et al.(2013)]{mcquilan13} McQuillan, A., 
Aigrain, S., \& Mazeh, T.\ 2013, \mnras, 432, 1203 

\bibitem[Meadows 
\& Crisp(1996)]{meadows96} Meadows, V.~S., \& Crisp, D.\ 1996, \jgr, 101, 4595 

\bibitem[Misra et al.(2014)]{misra14} Misra, A., Meadows, V., 
Claire, M., \& Crisp, D.\ 2014, Astrobiology, 14, 67 

\bibitem[Nielsen et 
al.(2013)]{nielsen13} Nielsen, M.~B., Gizon, L., Schunker, H., \& Karoff, C.\ 2013, \aap, 557, L10 

\bibitem[Pall{\'e} et al.(2009)]{palle09} Pall{\'e}, E., 
Zapatero Osorio, M.~R., Barrena, R., Monta{\~n}{\'e}s-Rodr{\'{\i}}guez, P., 
\& Mart{\'{\i}}n, E.~L.\ 2009, \nat, 459, 814 


\bibitem[Pasachoff et al.(2011)]{pasachoff2011} Pasachoff, J.~M., 
Schneider, G., \& Widemann, T.\ 2011, \aj, 141, 112 


\bibitem[Pont et al.(2008)]{pont08} Pont, F., Knutson, H., 
Gilliland, R.~L., Moutou, C., \& Charbonneau, D.\ 2008, \mnras, 385, 109 


\bibitem[Rauer et 
al.(2011)]{rauer11} Rauer, H., Gebauer, S., Paris, P.~V., et al.\ 2011, \aap, 529, A8 


\bibitem[Reid 
\& Hawley(2005)]{reid05} Reid, I.~N., \& Hawley, S.~L.\ 2005, New Light on Dark Stars Red Dwarfs, Low-Mass Stars, Brown Stars, by I.N.~Reid and S.L.~Hawley.~Springer-Praxis books in astrophysics and astronomy.~Praxis Publishing Ltd, 2005.~ ISBN 3-540-25124-3, 


\bibitem[Robinson et al.(2011)]{robinson11} Robinson, T.~D., 
Meadows, V.~S., Crisp, D., et al.\ 2011, Astrobiology, 11, 393 


\bibitem[Rothman et al.(2009)]{rothman09} Rothman, L.~S., Gordon, 
I.~E., Barbe, A., et al.\ 2009, \jqsrt, 110, 533 


\bibitem[Seager 
\& Sasselov(2000)]{seager00} Seager, S., \& Sasselov, D.~D.\ 2000, \apj, 537, 916 


\bibitem[Sidis 
\& Sari(2010)]{sidis10} Sidis, O., \& Sari, R.\ 2010, \apj, 720, 904 


\bibitem[Sioris et al.(2010)]{sioris10} Sioris, C.~E., Boone, 
C.~D., Bernath, P.~F., et al.\ 2010, Journal of Geophysical Research 
(Atmospheres), 115, D00L14


\bibitem[Smith 
\& Hunten(1990)]{smith90} Smith, G.~R., \& Hunten, D.~M.\ 1990, Reviews of Geophysics, 28, 117 


\bibitem[Snellen et al.(2013)]{snellen13} Snellen, I.~A.~G., de 
Kok, R.~J., le Poole, R., Brogi, M., \& Birkby, J.\ 2013, \apj, 764, 182 


\bibitem[Sosey et al.(2012)]{sosey12} Sosey, M., Hanley, C., 
Laidler, V., et al.\ 2012, Astronomical Data Analysis Software and Systems 
XXI, 461, 221 

\bibitem[Tyler et al.(1982)]{tyler82} Tyler, G.~L., Eshleman, 
V.~R., Anderson, J.~D., et al.\ 1982, Science, 215, 553 


\bibitem[van der Werf(2008)]{werf08} van der Werf, S.~Y.\ 
2008, \ao, 47, 153 


\bibitem[Vidal-Madjar et al.(2003)]{vidal03} Vidal-Madjar, A., 
Lecavelier des Etangs, A., D{\'e}sert, J.-M., et al.\ 2003, \nat, 422, 143 

\bibitem[Wilquet et al.(2012)]{wilquet12} Wilquet, V., Drummond, 
R., Mahieux, A., et al.\ 2012, \icarus, 217, 875 

\bibitem[Zombeck(1990)]{zombeck90} Zombeck, M.~V.\ 1990, Handbook of Space Astronomy and Astrophysics
Cambridge: University Press, 1990, 2nd ed.


\end{thebibliography}
\end{document}